\documentclass[fleqn,12pt,twoside]{article}
\usepackage{espcrc1}
\usepackage{graphicx}

\newcommand{\AmS}{{\protect\the\textfont2
  A\kern-.1667em\lower.5ex\hbox{M}\kern-.125emS}}

\title{The nuclear pairing problem: new perspectives}
\author{
Vladimir Zelevinsky\address{National Superconducting Cyclotron Laboratory and Department
of Physics and Astronomy, Michigan State University, East Lansing,
MI 48824-1321, USA}
\thanks{Supported by the NSF grants PHY-0070911 and
PHY-0244453.}
and Alexander Volya\address{Department of Physics, Florida State University,
Tallahassee, FL 32306-4350, USA}
\thanks{Supported by the USA Department of Energy
grant DE-FG02-92ER40750.}
}
\begin{document}
\maketitle

\begin{abstract}
The overview of the Exact Pairing technique based on the quasispin
symmetry is presented. Extensions of this method are discussed in
relation to mean field, quadrupole collectivity, electromagnetic
transitions, and many-body level density. Realistic calculations
compared with experimental data are used to support the methods as
well as to emphasize the manifestations of pairing correlations in
nuclear many-body systems.
\end{abstract}

\section{Introduction}

The pairing forces are known to be one of the main ingredients of
the residual interaction between the nucleons inside the nucleus.
Superconducting pairing correlations significantly influence all
nuclear properties. The role of pairing correlations increases for
weakly bound nuclei far from stability, and some nuclei turn out
to be bound only due to the pairing. However, the traditional
methods of solving the pairing problem borrowed from macroscopic
physics are insufficient for the correct description of the
pairing in nuclei and other finite mesoscopic systems. Below we
discuss the properties of the exact numerical solution of the
pairing problem based on the seniority classification. This
approach eliminates the conceptual drawbacks of the standard
schemes and opens new perspectives for theoretical advances. Below
we concentrate on developing new approximations for the many-body
problem in heavy nuclei where the full shell model diagonalization
is not feasible.

\section{Standard approach and its drawbacks}

The pairing \cite{BM} is a part of the general quantum many-body
problem. We consider this problem in the mean field framework.
Particles are moving in the initial ``bare" field that in general
has both bound orbitals and continuous spectrum. The continuum
will be considered elsewhere; here we assume a given set of
discrete single-particle levels $|jm)$ with bare energies
$\epsilon_{j}$. The levels are degenerate in quantum numbers $m$
because of rotational and time reversal invariance. The pairing
interaction,
\begin{equation}
H_{p}=\sum_{jj'}G_{j'j}P^{\dagger}_{j'}P_{j},     \label{1}
\end{equation}
transfers pairs of time-conjugate partners,
\begin{equation}
P_{j}=\frac{1}{2}\sum_{m}a_{jm}a_{j\tilde{m}},    \label{2}
\end{equation}
between the orbitals. The time-reversal orbital is, in spherical
basis, $|j\tilde{m})=(-)^{j-m}|j-m)$. An interaction of this type
is at the core of superconductivity in condensed matter \cite{BCS}
and of pairing correlations in nuclei \cite{BMP}.

The BCS theory of macroscopic superconductivity \cite{BCS}
generalizes the mean-field approach and constructs the ground
state as a self-consistent condensate of pairs with a certain
distribution of pairs over the levels (plane waves in uniform
superconductors) that minimizes the sum of the mean field energy
and pairing energy. The condensate does not have a certain number
of particles but the average number is controlled by the chemical
potential $\mu$, while fluctuations are small for macroscopic
media. The nuclear analog of the BCS approach was fully developed
by Belyaev \cite{Bel}; it was shown that pairing influences all
aspects of nuclear structure $-$ binding energy, odd-even effects,
occupation factors, transition and beta-decay probabilities,
quasiparticle excitations, collective modes, moments of inertia
and level density, see the recent review paper \cite{Dean}.
Reaction amplitudes and fission processes are also strongly
sensitive to pairing.

Belyaev also pointed out the shortcomings of the BCS approximation
for finite systems. The particle number violation can be critical
for small systems, especially on the edge of stability. There
exist vast literature devoted to the corresponding corrections
\cite{LN}. Another drawback is related to the mean field character
of the BCS solution. As a signature of the developed pairing, the
quasiparticle spectrum acquires the gap, $\epsilon_{j}\rightarrow
e_{j}=\sqrt{(\epsilon_{j}-\mu)^{2}+\Delta_{j}^{2}}$, which is
determined by the BCS equation
\begin{equation}
\Delta_{j}=\sum_{j'}G_{jj'}\,\frac{\Delta_{j'}}{2e_{j'}}.
                                        \label{3}
\end{equation}
In a large system with a continuous single-particle spectrum, eq.
(\ref{3}) always has a non-trivial solution (Cooper effect),
although at weak interaction the gap is exponentially small.
Contrary to that, in a finite system with a discrete spectrum
$\epsilon_{j}$, the non-zero solution appears only if typical
pairing matrix elements $G_{j'j}$ exceed a critical value of the
order of the single-particle level spacing. Otherwise, the gap
disappears revealing a sharp phase transition to the normal phase
as a function of the pairing strength. As seen in large-scale
shell model calculations, pairing correlations in a nucleus do not
vanish momentarily but smoothly decrease. Similarly, the thermal
phase transition is extended, and a long high-temperature tail of
pairing correlations was observed in shell model simulations
\cite{big} as well as in experiments with small superconducting
particles \cite{TS}. The HFB method \cite{Good}, making a step
beyond the BCS in the direction of accounting for the interplay
between the mean field and pairing effects, does not cure the
mentioned drawbacks.

\section{Exact solution of the pairing problem}

The exact algebraic solution of the pairing problem was developed
by Richardson in the series of papers, see \cite{rich} and
references therein. Unfortunately, this formalism turns out to be
numerically quite complicated and works only for specific choices
of the matrix elements $G_{j'j}$. We suggested \cite{EP} a direct
method of a simple numerical solution of the pairing problem based
on the quasispin algebra. The operators $P_{j}, P^{\dagger}_{j}$
and the occupancy of the level $j$,
$N_{j}=\sum_{m}a^{\dagger}_{jm}a_{jm}$, obey the commutation
relation of the $SU(2)$ algebra forming the partial quasispin
vector ${\bf L}_{j}$. This formalism is known long ago
\cite{racah,RT} to give the simple and exact solution for a
degenerate limit.

In a general many-level case, the pairing interaction does not
involve unpaired particles, and therefore their number on each
$j$-level, the partial seniority $s_{j}$, is preserved. For the
quasispin quantum number $\Lambda_{j}$ we have
\begin{equation}
{\bf L}_{j}^{2}=\Lambda_{j}(\Lambda_{j}+1), \quad
s_{j}=\frac{\Omega}{2}- 2\Lambda_{j},                 \label{4}
\end{equation}
where $\Omega_{j}$ is the capacity of the $j$-level ($2j+1$ in the
spherical shell model). A given set of partial seniorities
$s_{j}$, or quasispins $\Lambda_{j}$, defines a class of states
closed with respect to the pairing interaction, regardless of the
values of matrix elements $G_{jj'}$. The states with different
total spin $J$ within a class are degenerate in the absence of
other types of interaction. The ground state of an even-even
nucleus with $J=0$ belongs to the class with all $s_{j}=0$. The
dimensions of the classes are never exceedingly large, and, with a
sparse pairing Hamiltonian matrix, the numerical solution is very
fast.

Already at this stage, one can compare the results of the exact
solution with those of the BCS approximation for the same pairing
matrix elements and single-particle energies (long ago such a
comparison for simple examples on the existed level of accuracy
was considered \cite{Kerman} as an argument in favor of the BCS
approach). The detailed comparison for a typical spherical nucleus
$^{114}$Sn within the space of $g_{7/2}, d_{5/2}, d_{3/2},
s_{1/2}$, and $h_{11/2}$ neutron orbitals was made in Ref.
\cite{Yad}. This case, with the filled $g_{7/2}$ and $d_{5/2}$
levels, is not the most favorable for the BCS method.
Nevertheless, with 14 neutrons above the double-magic $^{100}$Sn,
the results of the BCS and exact solution are reasonably close.
While in the BCS theory the occupation numbers in the ground state
are expressed in terms of the coherence factors of the Bogoliubov
transformation,
\begin{equation}
N_{j}=v_{j}^{2}=1-u_{j}^{2},                   \label{5}
\end{equation}
in the exact solution one needs to distinguish between these three
numbers since the coherence factors are certain matrix elements
between the states $|N,s\rangle$ of different nuclei,
\begin{equation}
v_{j}(N)=\langle N-1,s_{j}=1|a_{j}|N,0\rangle, \quad u_{j}(N)=
\langle N+1,s_{j}=1|a^{\dagger}_{j}|N,0\rangle. \label{6}
\end{equation}
Typical differences for $^{114}$Sn are on the level of 10\%. The
effect is more pronounced for the pair transfer amplitudes
$\langle N+2,0|P^{\dagger}_{j}|n,0\rangle$ and $\langle N,0|P_{j}|
N-2,0\rangle$ which differ on the level of 20\%, whereas the BCS
result is around their mean value.

\begin{figure}
\begin{center}
\includegraphics[width=3.3 in]{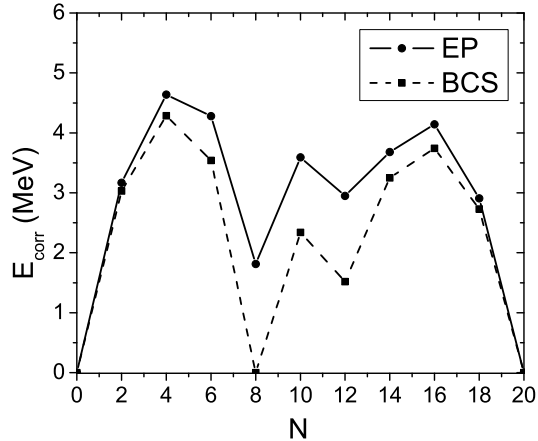}
%\vspace{-0.2 cm}
\end{center}
\caption{Pairing correlation energy along the chain of Ca isotopes.
\label{CaEcorr}}
%\vspace{-0.5 cm}
\end{figure}
%Fig. 1. Correlation energy for Ca isotopes.

As discussed in Ref. \cite{COMEX}, the situation is quite
different for nuclei as $^{48}$Ca, Fig.~\ref{CaEcorr}. Here the
BCS approximation results in a trivial solution because of the
complete filling of the $0f_{7/2}$ orbital. The exact solution
predicts a noticeable correlation energy close to 2 MeV. A
quenching of pairing effects is expected in the BCS theory for
$^{52}$Ca, where the orbital $1p_{3/2}$ is filled; the exact
solution again differs by almost 2 MeV. Thus, the BCS theory is
unreliable for many practically important cases, especially
outside of the valley of stability, where the structure and the
very existence of nuclei are very sensitive to magic numbers, and
the latter can be different from the region of stability being
determined by the details of the mean field and interaction that,
in turn, includes pairing.

The pairing solution in a given seniority sector gives, along with
the ground state, many excited states with the same quantum
numbers. In the class of $s=0$, all these states are built without
unpaired particles, and we can consider them as progressively
worsened copies of the pair condensate with a different pair
distribution over the orbitals. In the pure BCS approach, we have
only quasiparticle excitations. Adding the quasiparticle random
phase approximation as the next step beyond BCS, one obtains, see
\cite{RPA} and references therein, the so-called pair vibrations
with excitation energies $\omega$ as the roots of the secular
equation for the amplitudes $A_{j}$,
\begin{equation}
A_{j}=\sum_{j'}G_{jj'}\frac{2e_{j'}A_{j'}}{4e_{j'}^{2}-\omega^{2}}.
                                             \label{7}
\end{equation}
This equation has the solution $\omega=0$, when $A_{j}$ coincide
with the condensate amplitudes $\Delta_{j}$ of eq. (\ref{3}), plus
other roots above pair breaking threshold $2\Delta$. As excitation
energy grows, the abundant states found with the exact solution
have gradually decreasing collectivity that can be measured with
the off-diagonal matrix elements of the operators $N_{j}$. (It is
known that many nuclei have unexplained $0^{+}$ levels in the
region above the gap.) In the upper part of the spectrum one can
see signatures of chaotic mixing \cite{VZB} with no random
interactions included. Moreover, the exponential convergence of
the numerical results as a function of truncated dimension
\cite{ECM} confirms the onset of chaoticity at high level density
in the pure pairing problem. The further studies in this direction
are promising.

\section{On the way to the full interaction \label{sec::BE2}}

The exact solution for pairing (EP) is merely the first step in
the nuclear many-body problem. Even in the cases, such as Sn
isotopes, where pairing is the predominant part of the residual
interaction, one cannot ignore the presence of other parts. Since
our entire discussion assumes that the full shell model solution
in the required single-particle space is not feasible or too
time-consuming, we can try to develop intermediate approximations
based on the exact pairing solution as a starting point.

First of all, the monopole part (a combination of all
particle-particle interactions, except for exactly treated
pairing, transformed to the particle-hole channel) is necessary in
order to account for the smooth change of the mean field along
with the change of occupation. This ``EP + monopole" approximation
works well for global quantities, especially if we are interested
in the evolution of those quantities with the mass number. In
fact, this part was taken into account in the calculation of
correlational energy in Fig. 1.

At the next step, we move to the analog of the HFB method, where
the pairing and the mean field, including the deformation effects,
become mutually interrelated. The EP solution determines the
occupation numbers $N_{j}$ in the ground state. Similar to what is
used at nonzero temperature, we build the Hartree-Fock
approximation on those occupation numbers instead of the Slater
determinant. In the new mean field (possibly, deformed) we can
solve a new the pairing problem and repeat these cycles up to
convergence. Fig. 2 shows the neutron separation energies for the
longest chain of tin isotopes calculated in this approximation
\cite{VBZPTP}. A popular model ``pairing + quadrupole-quadrupole
interaction" was analyzed from this viewpoint in Ref.
\cite{Volya}. The model is usually treated as an interpolation
picking up the most important interaction parts in both channels.
Being considered literally, this model reveals strong pairing
effects coming from the usually ignored exchange terms from the
quadrupole interaction \cite{Yad}.

\begin{figure}
\begin{center}
\includegraphics[width=3.3 in]{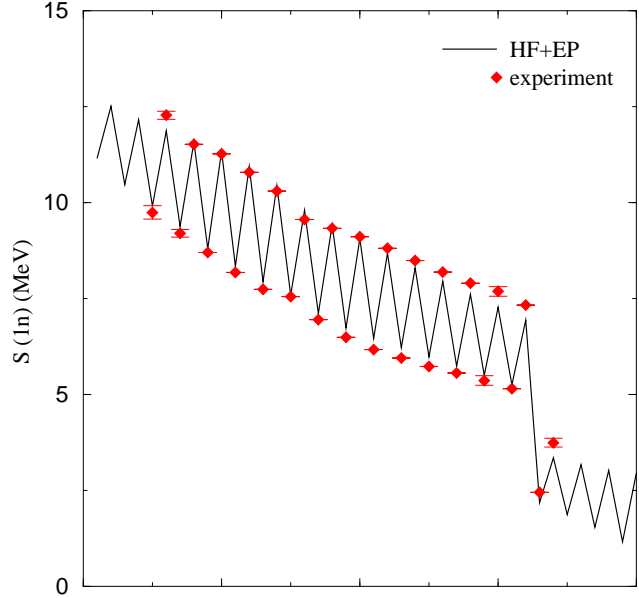}
%\vspace{-0.2 cm}
\end{center}
\caption{One-neutron separation energies for Sn isotopes,
including those beyond $^{132}$Sn. Single-particle energies were
obtained from the fully self-consistent spherically symmetric
solution of HF equations, using the SKX interaction
\cite{brown98}, with the EP solution based on the renormalized
G-matrix interaction from Ref. \cite{jensen}. \label{Sn_HFEP}}
%\vspace{-0.2 cm}
\end{figure}

\begin{figure}
\begin{center}
\includegraphics[width=6 in]{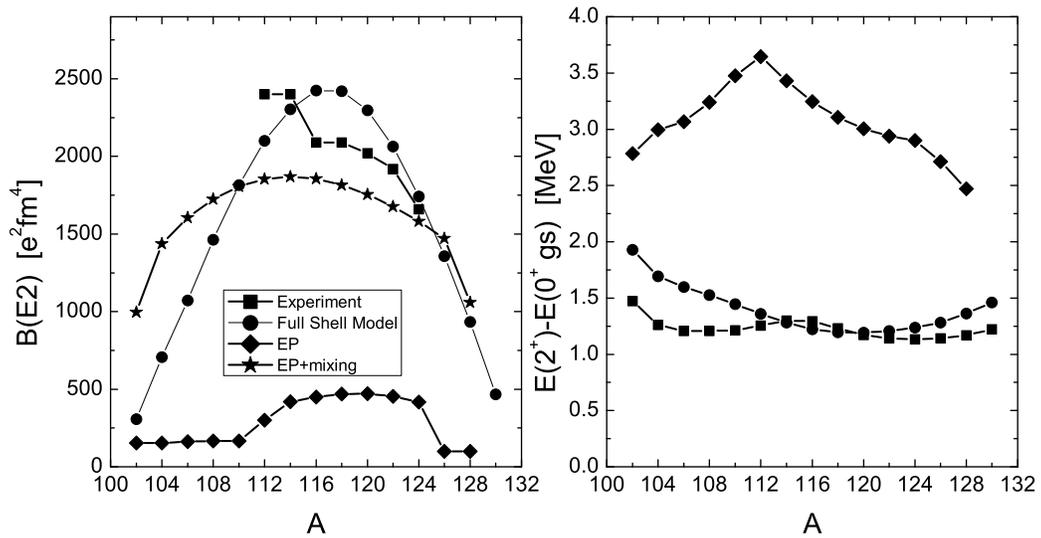}
\vspace{-0.4 cm}
\end{center}
\caption{The transition probabilities $B$(E2;$2^{+}_{1}\rightarrow
0^{+}_{1})$, see the text. \label{BE2E2}}
\end{figure}

Currently, there exists a strong interest to the behavior of
electromagnetic transition probabilities
$B$(E2;$2^{+}_{1}\rightarrow 0^{+}_{1})$ from the first excited
quadrupole states to the ground state in the chain of even-even
tin isotopes \cite{Giamb}. The $2^{+}_{1}$ energies are known to
be approximately constant for all non-magic isotopes from
$^{102}$Sn to $^{130}$Sn, Fig \ref{BE2E2}, right panel. This can
be considered as absence of considerable deformation effects
usually attributed to the interaction between valence neutrons and
protons. Meanwhile, the experimental $B$(E2) values measured for
$A\geq 112$ display a non-trivial $A$-dependence shown on the left
panel in Fig. 3.
The comparison of different types of calculations shown in Fig.
\ref{BE2E2} reveals the important role of collective correlations
beyond pairing. EP + monopole calculations, although quite
accurate for the ground state and lowest pair vibration states,
are insufficient once there are unpaired particles in the system;
the residual interactions beyond pairing become crucial. In Fig.
\ref{BE2E2}, the full shell model calculation agrees reasonably
well with available data for both $B$(E2;$2^{+}_{1}\rightarrow
0^{+}_{1})$ and energy of the lowest $2^+$ state. The exact EP +
monopole calculation that ignores all other interactions still
reproduces the total ground state energy within 0.4 MeV error; but
the results are poor for the $2^+$ excited state that is treated
as a seniority $s=2$ state. Indeed, gross underestimates of
$B$(E2;$2^{+}_{1}\rightarrow 0^{+}_{1})$, and $E(2^+)$ being
almost a factor of 2 high indicate a substantial lack of
collectivity. The quadrupole collectivity increases significantly
when states with unpaired particles are allowed to interact
breaking seniority. The calculation of $B$(E2) values for the most
quadrupole-coherent superposition of seniority $s=2$ states is
shown on left panel in Fig. \ref{BE2E2} with stars. The remaining
discrepancy supposedly comes from the admixture of states with
$s>2$ and from breaking the proton closed shell. In this
situation, the results of the planned at the NSCL experiment for
measuring $B$(E2) for lighter Sn isotopes are of significant
interest.

\section{Density of states in paired systems}

\begin{figure}
\begin{center}
\includegraphics[width=4.5 in]{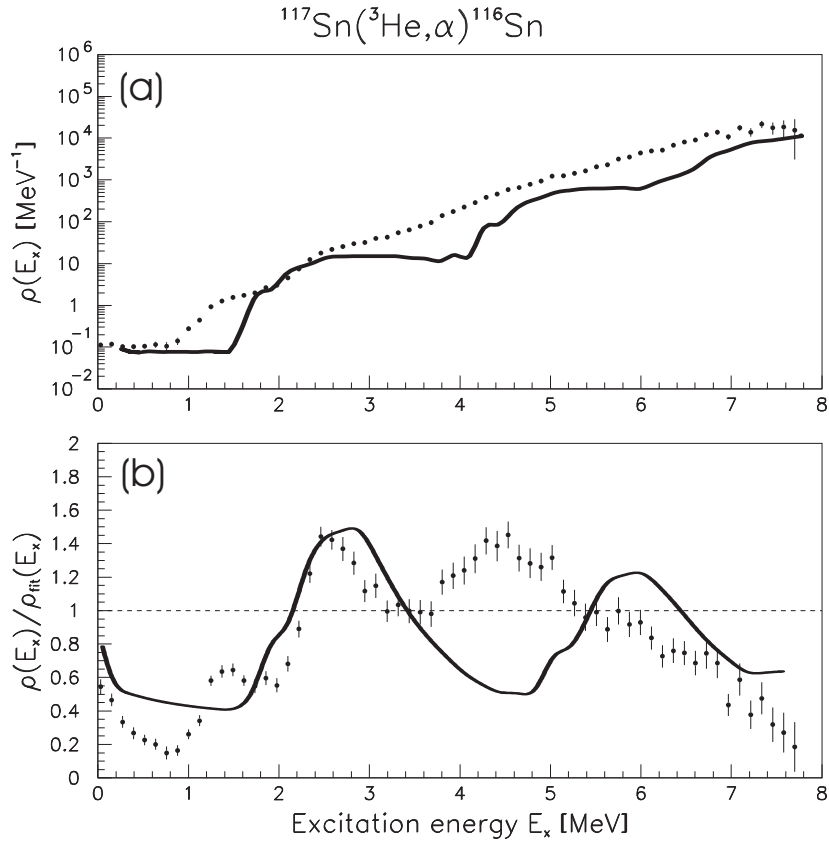}
%\vspace{-0.2 cm}
\end{center}
\caption{Density of states measured in $^{116}$Sn, courtesy of U.
Agvaanluvsan. The solid curves show the density of states computed
for the same system with just pairing+monopole interactions. In
panel (a) full density of states from the experiment is compared
to the theoretical calculation. To emphasize the oscillations,
panel (b) displays the same comparison with the level density
plotted as a ratio to the fitted curve defined as $\rho_{\rm
fit}(E)=A \exp(B E)$ with $A$ and $B$ being parameters of the fit.
For theoretical curve the amplitude of oscillations is scaled down by a
factor of 1/2.5. \label{sn116ro}}
\end{figure}

Pairing correlations leave their footprints far beyond few
low-lying states. Recently developed new experimental techniques
allow for determination of the level density \cite{melby99} up to
neutron separation energy. The experimental density of states in
$^{116}$Sn is shown in Fig. \ref{sn116ro}, where the solid line
shows the density of states computed with the EP+monopole
technique. The calculated curve is significantly lower than the
experiment. The absent in this approximation interactions of
unpaired particles broaden the density of states \cite{big} and
bring specific high-lying states lower. The same phenomenon is
responsible for the lowering of the $2^+_{1}$ state discussed in
Sec. \ref{sec::BE2}.

Despite poor general agreement between the data and calculations
in the EP+monopole approximation, the full density of states
exhibits bumps that can be traced to the prominent seniority peaks
obtained in the model. In the calculations with pure pairing, and
large degeneracy within a given seniority class, the opening of a
new class with higher seniority brings in a large enhancement of
the level density. The peaks have to be considerably, or maybe
even completely, washed out by other types of interactions lifting
degeneracies. This can be noted by examining Fig.
\ref{sn116ro}(b), where presence of peaks is clear in both
experiment and theory, but in order to fit the scale of the peaks
observed in experiment a reduction factor of 2.5 is needed for the 
theoretical curve.

\begin{figure}
\begin{center}
\includegraphics[width=4 in]{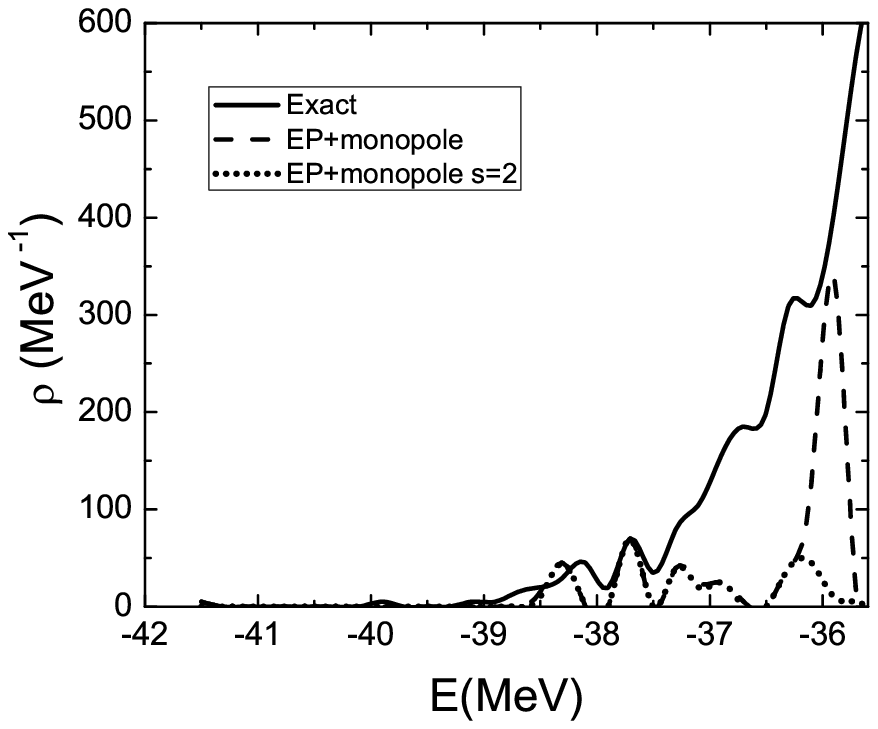}
\vspace{-0.5 cm}
\end{center}
\caption{Density of positive parity states calculated for
$^{106}$Sn.: full shell model calculation, solid line; EP+monopole
calculation, dashed line; the contribution from seniority $s=2$
states in the EP+monopole calculation. \label{dEPM106p}}
\vspace{-0.7 cm}
\end{figure}

The role of pair breaking in creating peaks in density of states
can be seen in Fig. \ref{dEPM106p} for the case of $^{106}$Sn.
Although the complete experimental data are not available for this
system, the relatively small valence space allows here for the
exact shell model diagonalization shown with the solid curve. The
EP+monopole calculations, dashed line, exhibit large peaks,
similar to those in Fig. \ref{sn116ro}. The peaks coming from
pairing can be still well recognized in the full calculation with
all interactions included. To see the relation of these peaks to
seniority we show the density of $s=2$ states with the dotted
line. The dashed and dotted curves coincide at low energies, but
seniority $s=2$ subspace gets exhausted by the time energy reaches
-36 MeV (this is relative to the $^{100}$Sn core and equivalent to
about 6 MeV excitation energy). Then another pair has to be
broken, and a large peak in the dashed line corresponding to $s=4$
follows, while the dotted $s=2$ line goes to zero. The related
discussions, other experimental results, and model studies can be
found in the review \cite{Dean}. We have to mention recent
theoretical suggestions for calculating the shell model level
density avoiding the full diagonalization \cite{levden}.
\section{Conclusions}
The crucial importance of pairing correlations in nuclear
structure has been identified more than half a century ago, and it
is remarkable that today this topic is as fresh as ever before.
New theoretical and experimental techniques allow us to uncover
yet more and more bright manifestations of these correlations.
This paper is centered around just one of the new theoretical
methods. Using algebraic simplicity of the pairing interaction, we
can drastically extend the applicability of the nuclear shell
model and eliminate all shortcomings of the conventional treatment
borrowed from macroscopic physics. The same approach should be
useful for other mesoscopic Fermi-systems with superfluid or
superconducting correlations.

The interactions beyond pairing can be incorporated in the
perturbative  manner which would allow one to study transitions
from small systems to superconducting media. We discussed the role
of pairing and other interactions in electromagnetic transitions,
and quadrupole collectivity. The ability of the EP method to
generate all many-body states was crucial. Our study of density of
states reaffirmed that even at relatively high excitation energy,
and with all non-pairing interactions present, pairing
correlations are still an important part of the nuclear many-body
dynamics. Along with the realization of vibrational and rotational
states on top of the EP, as well as of details of the quenched
phase transition to a normal state, where we suggested invariant
correlational entropy as a signature of phase transformations
\cite{ICE}, this should be a subject of future research. We did
not discuss here the continuum effects and the bridge to the
reaction theory, where the role of pairing is also crucial
\cite{PRC67,new}.

\section{Acknowledgements}

Collaboration with B.A. Brown and M. Horoi is gratefully
acknowledged; we thank U. Agvaanluvsan for making the data on the
level density accessible prior to publication, T. D\o ssing, R.
Johnson, A. Schiller and K. Starosta for motivating discussions.

\end{document}